\documentstyle[12pt]{article}
\begin{document}
\def\beq{\begin{equation}}
\def\eeq{\end{equation}}
\def\bea{\begin{eqnarray}}
\def\eea{\end{eqnarray}}
\def\ve{\vert}
\def\vel{\left|}
\def\ver{\right|}
\def\nnb{\nonumber}
\def\ga{\left(}
\def\dr{\right)}
\def\aga{\left\{}
\def\adr{\right\}}
\def\rar{\rightarrow}
\def\nnb{\nonumber}
\def\la{\langle}
\def\ra{\rangle}
\def\lla{\left<}
\def\rra{\right>}
\def\ba{\begin{array}}
\def\ea{\end{array}}
\def\tep{$B \rar K \ell^+ \ell^-$}
\def\tepm{$B \rar K \mu^+ \mu^-$}
\def\tept{$B \rar K \tau^+ \tau^-$}
\def\ds{\displaystyle}


\def\lesssim{\mathrel{\mathpalette\vereq<}}
\def\vereq#1#2{\lower3pt\vbox{\baselineskip1.5pt \lineskip1.5pt
\ialign{$\m@th#1\hfill##\hfil$\crcr#2\crcr\sim\crcr}}}

\def\gtrsim{\mathrel{\mathpalette\vereq>}}

\def\alt{\lesssim}
\def\agt{\gtrsim}




\def\bos{\lower 0.5cm\hbox{{\vrule width 0pt height 1.2cm}}}
\def\boss{\lower 0.35cm\hbox{{\vrule width 0pt height 1.cm}}}
\def\aaa{\lower 0.cm\hbox{{\vrule width 0pt height .7cm}}}
\def\dol{\lower 0.4cm\hbox{{\vrule width 0pt height .5cm}}}


\title{ {\Large {\bf 
Lepton polarization and CP--violating effects in 
$B \rar K^\ast \tau^+ \tau^-$ decay
in standard and two Higgs doublet models} } }

\author{\vspace{1cm}\\
{\small T. M. Aliev \thanks
{e-mail: taliev@metu.edu.tr}\,\,,
M. Savc{\i} \thanks
{e-mail: savci@metu.edu.tr}} \\
{\small Physics Department, Middle East Technical University} \\
{\small 06531 Ankara, Turkey} }
\date{}

\begin{titlepage}
\maketitle
\thispagestyle{empty}

\begin{abstract}
\baselineskip  0.7cm
The most general model independent expressions for the CP--violating 
asymmetry, longitudinal,
transversal and normal polarizations of leptons are derived. Application of these
general results to the concrete models such as Standard model and three
different types of two Higgs doublet model is discussed. 
\end{abstract}


\end{titlepage}

\section{Introduction}
Rare $B$ meson decays, induced by flavor--changing neutral current (FCNC) 
$b \rar s$ transitions, is one of the most promising research area in high
energy physics. Theoretical interest to the rare $B$ decays lies in their
role as a potential precision testing ground for the standard model (SM) 
at loop level and experimentally these decays will
provide quantitative information about the Cabibbo--Kobayashi--Maskawa 
(CKM) matrix elements $V_{td},~V_{ts}$ and $V_{tb}$. Besides, these rare
decays have the potential of establishing new physics beyond SM, such as two
Higgs doublet model (2HDM), minimal supersymmetric extension of the SM
(MSSM), left--right models, etc.

At present, the main interest is focused on the rare $B$ meson decays, for
which the SM predicts "large" branching ratios and that can be potentially
measurable at working $B$ factories and LHC. The rare 
$B \rar K^\ast \ell^+ \ell^-$ $(\ell=e,~\mu,~\tau)$ decays are such ones.
At the same time this decay constitutes a suitable tool in looking for new
physics beyond the SM.
At quark level the process is described by the $b \rar s \ell^+ \ell^-$     
transition. This transition in framework of the SM and its various
extensions have been extensively investigated \cite{R1}--\cite{R15}. 

One efficient way in establishing new physics is the measurement of lepton
polarization. This problem is widely discussed in literature for the
$b \rar s \ell^+ \ell^-$ decay \cite{R16}--\cite{R19}. 
Note that all previous studies for the lepton polarization, except 
the work \cite{R19}, have been limited to SM and its minimal extensions. 
In \cite{R19} the analysis of the $\tau$ lepton polarization for the 
$b \rar s \tau^+ \tau^-$ decay was performed in a model independent way.    
In this work twelve (ten local and two nonlocal) four--Fermi operator interactions 
were introduced in a model independent way instead of three independent
structures which are present in SM. 

It is well known that the theoretical analysis of the inclusive decays is
rather easy but their experimental detection is quite difficult. However for
exclusive decays the situation is contrary to this case, i.e., their
experimental study is easy but theoretical investigation is difficult. This
is due to the fact that the description of the exclusive decays requires
form factors, i.e., the matrix elements of the effective Hamiltonian between
initial $B$ and final meson states. This problem is related to to the
nonperturbative sector of the QCD and it can be solved only by means of a
nonperturbative approach. 

These matrix elements have been investigated in framework of different
approaches such as chiral theory \cite{R20}, three point QCD sum rules
method \cite{R21}, relativistic quark model by the light--front formalism 
\cite{R22}, effective heavy quark theory \cite{R23} and light cone QCD sum
rules \cite{R24,R25}. 

The aim of the present paper is to perform a comprehensive study the lepton 
polarizations and CP--violating asymmetry in the
exclusive $B \rar K^\ast \ell^+ \ell^-$ $(\ell = \mu,~\tau)$ decay in the SM
and three versions of the 2HDM. Note that this decay in framework of 2HDM
(model I and model II) were investigated in \cite{R15} (second reference),
where $P_L$ and $P_T$ and $A_{FB}$ were studied. Here in this work we extend
these considerations by studying in model III, including normal polarization
of $\tau$ lepton and CP--violating asymmetry, using the current
limits for the parameters of 2HDM coming from low energy experiments like 
$B^0$--$\bar B^0$ mixing, $\rho$--parameter analysis and $b \rar s \gamma$
decay.  

The paper is organized as follows. In section 2, starting from a general form of
four--Fermi interactions we derive model independent expressions for the
longitudinal, transversal and normal polarizations. In section 3, we apply
the above--mentioned general results of the lepton polarizations
to SM and to three types of 2HDM 
(so called models I, II and III). A brief summary of our results is
presented in this section.

\section{Lepton polarizations}

In this section we present the expressions for the longitudinal, transversal
and normal polarizations of the $\tau$ lepton in a model independent way.
For this aim we follow \cite{R19} where the matrix element for 
$b \rar s \tau^+ \tau^-$ transition is given in terms of twelve most general
independent four--Fermi interactions:
\bea
{\cal M} &=& \frac{G\alpha}{\sqrt{2} \pi}
 V_{tb}V_{ts}^\ast
\Bigg\{ C_{SL} \, \bar s i \sigma_{\mu\nu} \frac{q^\nu}{q^2} (m_s L) b
\, \bar \ell \gamma_\mu \ell + C_{BR}\, \bar s i \sigma_{\mu\nu}
\frac{q^\nu}{q^2} (m_b R) b \, \bar \ell \gamma_\mu \ell \nnb \\
&+&C_{LL}\, \bar s_L \gamma_\mu b_L \,\bar \ell_L \gamma^\mu \ell_L +
C_{LR} \,\bar s_L \gamma_\mu b_L \, \bar \ell_R \gamma^\mu \ell_R +
C_{RL} \,\bar s_R \gamma_\mu b_R \,\bar \ell_L \gamma^\mu \ell_L \nnb \\
&+&C_{RR} \,\bar s_R \gamma_\mu b_R \, \bar \ell_R \gamma^\mu \ell_R +
C_{LRLR} \, \bar s_L b_R \,\bar \ell_L \ell_R +    
C_{RLLR} \,\bar s_R b_L \,\bar \ell_L \ell_R \nnb \\         
&+&C_{LRRL} \,\bar s_L b_R \,\bar \ell_R \ell_L +
C_{RLRL} \,\bar s_R b_L \,\bar \ell_R \ell_L+                        
C_T\, \bar s \sigma_{\mu\nu} b \,\bar \ell \sigma^{\mu\nu}\ell \nnb \\                             
&+&i C_{TE}\,\epsilon^{\mu\nu\alpha\beta} \bar s \sigma_{\mu\nu} b \, 
\bar \ell \sigma_{\alpha\beta} \ell  \Bigg\}~,
\eea
where $C_{XX}$ are the coefficient of the four--Fermi interactions.
Among them, there are two non--local Fermi interactions denoted by $C_{SL}$
and $C_{BR}$, which correspond to $-2C_7^{eff}$ in the SM. Two ($C_{LL}$,~ $C_{LR}$)
of the four vector type interactions ($C_{LL}$, $C_{LR}$, $C_{RL}$ and
$C_{RR}$) are also present in the SM (in the forms of $C_{LL}=C_9^{eff} - C_{10}$
and $C_{LR}=C_9^{eff} + C_{10}$). $C_{LRLR}$, $C_{RLLR}$, $C_{LRRL}$ and
$C_{RLRL}$ are the scalar type interactions and the last two terms with
coefficients $C_T$ and $C_{TE}$ are the tensor type interactions.

For simplicity we will take the mass of the strange quark to be zero and
neglect tensor type interactions, since it was indicated in \cite{R26}
that the physical observables are not sensitive to the presence of the tensor type
interactions. Having established the matrix element for the
$b \rar s \tau^+ \tau^-$ transition, our next problem is calculation of
the matrix elements, 
$\lla K^\ast \vel \bar s \gamma_\mu (1\pm \gamma_5) b \ver B \rra$, 
$\lla K^\ast \vel \bar s i \sigma_{\mu\nu} q^\nu (1+ \gamma_5) b \ver B \rra$
and $\lla K^\ast \vel \bar s (1\pm \gamma_5) b \ver B \rra$, in order to be
able to calculate the physically measurable quantities  at hadronic level.
These matrix elements
can be written in terms of the form factors in the following way:
\bea
\lefteqn{
\lla K^\ast(p_{K^\ast},\varepsilon) \vel \bar s \gamma_\mu (1 \pm \gamma_5) b \ver
B(p_B) \rra =} \nnb \\
&&- \epsilon_{\mu\nu\rho\sigma} \varepsilon^{\ast\nu} p_{K^\ast}^\rho q^\sigma
\frac{2 V(q^2)}{m_B+m_{K^\ast}} \pm i \varepsilon_\mu^\ast (m_B+m_{K^\ast})
A_1(q^2) \mp i (p_B + p_{K^\ast})_\mu (\varepsilon^\ast q)
\frac{A_2(q^2)}{m_B+m_{K^\ast}} \nnb \\
&&\mp i q_\mu \frac{2 m_{K^\ast}}{q^2} (\varepsilon^\ast q)
\left[A_3(q^2)-A_0(q^2)\right]~, \nnb \\ \\
\lefteqn{
\lla K^\ast(p_{K^\ast},\varepsilon) \vel \bar s i \sigma_{\mu\nu} q^\nu 
(1 + \gamma_5) b \ver B(p_B) \rra =} \nnb \\
&&4 \epsilon_{\mu\nu\rho\sigma} \varepsilon^{\ast\nu} p_{K^\ast}^\rho q^\sigma
T_1(q^2) + 2 i \left[ \varepsilon_\mu^\ast (m_B^2-m_{K^\ast}^2) -
(p_B + p_{K^\ast})_\mu (\varepsilon^\ast q) \right] T_2(q^2) \nnb \\
&&+ 2 i (\varepsilon^\ast q) \left[ q_\mu -
(p_B + p_{K^\ast})_\mu \frac{q^2}{m_B^2-m_{K^\ast}^2} \right] T_3(q^2)~,
\eea
where $\varepsilon$ is the polarization vector of $K^\ast$ meson and
$q=p_B-p_{K^\ast}$ is the momentum transfer.
In order to ensure finiteness of (2) at $q^2=0$, we assume that
$A_3(q^2=0) = A_0(q^2=0)$. 
To calculate the matrix element
$\lla K^\ast \vel \bar s (1 \pm \gamma_5) b \ver B \rra$, we
multiply both sides of Eq. (2) by $q_\mu$ and use the equation of motion.
Neglecting the mass of the strange quark, we get
\bea
\lefteqn{
\lla K^\ast(p_{K^\ast},\varepsilon) \vel \bar s (1 \pm \gamma_5) b \ver
B(p_B) \rra =
\frac{1}{m_b} \Big\{ \mp i (\varepsilon^\ast q) (m_B+m_{K^\ast})
A_1(q^2)}\nnb \\
&&~~~~~~~~~~~~~\pm i (m_B-m_{K^\ast}) (\varepsilon^\ast q) A_2(q^2)
\pm 2 i m_{K^\ast} (\varepsilon^\ast q) \left[A_3(q^2)-A_0(q^2)\right]~\Big\}.
~~~~~~~~~~~
\eea
Using the equation of motion, the form factor $A_3$ can be expressed as a
linear combination of the form factors $A_1$ and $A_2$ (see \cite{R21})
\bea
A_3(q^2) = \frac{m_B+m_{K^\ast}}{2 m_{K^\ast}} A_1(q^2) -
\frac{m_B-m_{K^\ast}}{2 m_{K^\ast}} A_2(q^2)~.
\eea
Using this relation we obtain
\bea
\lla K^\ast(p_{K^\ast},\varepsilon) \vel \bar s (1 \pm \gamma_5) b \ver B(p_B) \rra
= \frac{1}{m_b} \Big\{\mp 2 i m_{K^\ast} (\varepsilon^\ast q) A_0(q^2)
\Big\}~.
\eea
From Eqs. (1), (2), (3) and (6) we get the following expression for the
matrix element of the $B \rar K^\ast \ell^+ \ell^-$ decay
\bea
\lefteqn{             
{\cal M} =
\frac{G \alpha}{4 \sqrt{2} \pi} V_{tb} V_{ts}^\ast }\nnb \\ 
&&\times \Bigg\{
\bar \ell \gamma_\mu(1-\gamma_5) \ell \, \Big[
-\epsilon_{\mu\nu\rho\sigma} \varepsilon^{\ast\nu} p_{K^\ast}^\rho q^\sigma
(2 A_1) -i\varepsilon_\mu^\ast B_1
+ i (\varepsilon^\ast q) (p_B+p_{K^\ast})_\mu
B_2 + i q_\mu (\varepsilon^\ast q) B_3 \Big] \nnb \\
&&+ \bar \ell \gamma_\mu(1+\gamma_5) \ell \, \Big[ 
-\epsilon_{\mu\nu\rho\sigma} \varepsilon^{\ast\nu} p_{K^\ast}^\rho q^\sigma
(2 C_1) - i \varepsilon_\mu^\ast D_1 + i (\varepsilon^\ast q)
(p_B+p_{K^\ast})_\mu D_2 + i q_\mu (\varepsilon^\ast q) D_3 \Big] \nnb \\
&&+\bar \ell (1-\gamma_5) \ell \, i (\varepsilon^\ast q) B_4 
+ \bar \ell (1+\gamma_5) \ell \, i (\varepsilon^\ast q) B_5 \Bigg\}~,
\eea
where
\bea
A_1 &=& (C_{LL} + C_{RL}) \frac{V(q^2)}{m_B+m_{K^\ast}} -
2 C_{BR} \frac{m_b}{q^2} \, T_1 , \nnb \\
B_1 &=& (C_{LL} - C_{RL}) (m_B+m_{K^\ast}) A_1 - 2 C_{BR} 
\frac{m_b}{q^2} (m_B^2-m_{K^\ast}^2) \, T_2~, \nnb \\ 
B_2 &=& \frac{C_{LL} - C_{RL}}{m_B+m_{K^\ast}} A_2 - 2 C_{BR} 
\frac{m_b}{q^2}  \left[ T_2 + \frac{q^2}{m_B^2-m_{K^\ast}^2} T_3 \right]~,
\nnb \\ 
B_3 &=& (C_{LL} - C_{RL}) \frac{2 m_{K^\ast}}{q^2} (A_3-A_0)-
2 C_{BR} \frac{m_b}{q^2} T_3~, \nnb \\
C_1 &=& A_1 ( C_{LL} \rar C_{LR}~,~~C_{RL} \rar C_{RR})~,\nnb \\ 
D_1 &=& B_1 ( C_{LL} \rar C_{LR}~,~~C_{RL} \rar C_{RR})~,\nnb \\
D_2 &=& B_2 ( C_{LL} \rar C_{LR}~,~~C_{RL} \rar C_{RR})~,\nnb \\
D_3 &=& B_3 ( C_{LL} \rar C_{LR}~,~~C_{RL} \rar C_{RR})~,\nnb \\
B_4 &=& - ( C_{LRRL} - C_{RLRL}) \ga \frac{2 m_{K^\ast}}{m_b} A_0 \dr~,\nnb \\
B_4 &=& - ( C_{LRLR} - C_{RLLR}) \ga \frac{2 m_{K^\ast}}{m_b} A_0 \dr~,\nnb \\
\eea
At this point we would like to make the following comment. 
The main difference from the SM case is, we have six different structures (after setting
$m_s=0$ and neglecting tensor interactions) in the inclusive channel, 
while under the same conditions we have only two new structures, namely scalar
type interactions proportional to $B_4$ and $B_5$. On the other hand there
appears no any new structure in the 2HDM. 

Having established this matrix element, let us now consider the final lepton
polarization. We define the following three orthogonal unit vectors:
\bea
\vec{e}_L &=& \frac{\vec{p}_1}{\ve \vec{p}_1 \ve}~, \nnb \\
\vec{e}_N &=& \frac{\vec{p}_{K^\ast}\times\vec{p}_1}
                   {\ve \vec{p}_{K^\ast}\times\vec{p}_1 \ve}~, \nnb \\
\vec{e}_T &=& \vec{e}_L\times\vec{e}_N~,
\eea
where $\vec{p}_1$ and $\vec{p}_{K^\ast}$ are the three--momenta of the
lepton $\ell^-$ and $K^\ast$ meson, respectively, in the center of mass of
final leptons. The differential decay rate for any spin direction $\vec{n}$
of the $\ell^-$ lepton, where $\vec{n}$ is a unit vector in the $\ell^-$
rest frame, can be expressed in the following form
\bea
\frac{d \Gamma(\vec{n})}{dq^2} = \frac{1}{2} \ga \frac{d \Gamma}{dq^2}\dr_0
\Big[1 + \Big( P_L \vec{e}_L + P_N \vec{e}_N + P_T \vec{e}_T \Big) \cdot
\vec{n}\Big]~,
\eea 
where the subscript "0" corresponds to the unpolarized differential decay
rate whose explicit form will be presented below. $P_L$, $P_N$ and $P_T$ are 
recognized as the longitudinal, normal and transversal polarizations, respectively.  
It follows from the definition of unit vectors $\vec{e}_i$ that $P_T$
obviously lies in the decay plane whose orientation is determined by the
vectors $\vec{p}_1$ and $\vec{p}_{K^\ast}$ and $P_N$ is perpendicular to
this plane. 

The expression for the unpolarized differential decay rate in Eq. (10) can be
written as:
\bea
\ga \frac{d \Gamma}{dq^2}\dr_0 &=& \frac{G^2 \alpha^2}{2^{14} \pi^5 m_B} 
\vel V_{tb} V_{ts}^\ast \ver^2 \lambda^{1/2} v \nnb \\
&\times& \Bigg\{
32 \lambda m_B^4 \Big[ \frac{1}{3} (m_B^2 s - m_\ell^2)
(\vel  A_1 \ver^2 + \vel  C_1 \ver^2 ) + 2 m_\ell^2 \, \mbox{\rm Re} 
(A_1 C_1^\ast)\Big] \nnb \\
&+& 96 m_\ell^2 \, \mbox{\rm Re} (B_1 D_1^\ast)-
\frac{4}{r} m_B^2 m_\ell \lambda \,
\mbox{\rm Re} [(B_1 - D_1) (B_4^\ast - B_5^\ast)] \nnb \\
&+&\frac{8}{r} m_B^2 m_\ell^2 \lambda \, \Big\{
\mbox{\rm Re} [(B_3^\ast + D_2^\ast - D_3^\ast) B_1] +
\mbox{\rm Re} [ (B_2^\ast - B_3^\ast + D_3^\ast) D_1] -
\mbox{\rm Re}(B_4 B_5^\ast) \Big] \Big\}~~~~~~~~ \nnb \\
&+&\frac{4}{r} m_B^4 m_\ell (1-r) \lambda \,
\Big\{\mbox{\rm Re} [(B_2 - D_2) (B_4^\ast - B_5^\ast)]\Big\} \nnb \\
&+& \frac{8}{r} m_B^4 m_\ell^2 (1-r) \lambda \,
\Big\{\mbox{\rm Re} [-(B_2 - D_2) (B_3^\ast - D_3^\ast)]\Big\}\nnb \\
&-& \frac{8}{r}m_B^4 m_\ell^2 \lambda (2+2 r-s)\, \mbox{\rm Re} (B_2 D_2^\ast)
-\frac{4}{r} m_B^4 m_\ell s \lambda \,
\mbox{\rm Re} [(B_3 - D_3) (B_4^\ast - B_5^\ast)] \nnb \\
&-&\frac{4}{r} m_B^4 m_\ell^2 s \lambda \,
\Big[ \vel B_3 \ver^2 + \vel D_3 \ver^2 - 
2 \, \mbox{\rm Re}(B_3 D_3^\ast)\Big]
+\frac{2}{r} m_B^2 (m_B^2-2 m_\ell^2) \lambda \,
\Big[ \vel B_4 \ver^2 + \vel B_5 \ver^2 \Big] \nnb \\
&-&\frac{8}{3rs} m_B^2 \lambda \,
\Big[m_\ell^2 (2-2 r+s)+m_B^2 s (1-r-s) \Big]
\Big[\mbox{\rm Re}(B_1 B_2^\ast) + \mbox{\rm Re}(D_1 D_2^\ast)\Big] \nnb \\ 
&+&\frac{4}{rs}\,
\Big[2 m_\ell^2 (\lambda-6 rs)+m_B^2 s (\lambda+12 rs) \Big]
\Big[ \vel B_1 \ver^2 + \vel D_1 \ver^2 \Big] \nnb \\
&+&\frac{4}{3rs} m_B^4 \lambda\,
\Big\{ m_B^2 s \lambda + m_\ell^2 \Big[ 2 \lambda + 3 s (2+2 r - s) \Big] \Big\}
\Big[ \vel B_2 \ver^2 + \vel D_2 \ver^2 \Big] 
\Bigg\}~.
\eea
The polarizations $P_L$, $P_N$ and $P_T$ are defined as:
\bea
P_i (q^2) = \frac{\ds{\frac{d \Gamma}{dq^2} (\vec{n}=\vec{e}_i) -
                      \frac{d \Gamma}{dq^2} (\vec{n}=-\vec{e}_i)}} 
                 {\ds{\frac{d \Gamma}{dq^2} (\vec{n}=\vec{e}_i) +
                      \frac{d \Gamma}{dq^2} (\vec{n}=-\vec{e}_i)}}~.
\eea
After lengthy calculations we get the following general expressions
for the longitudinal, transversal and normal
polarizations of the $\ell^-$ lepton (for $m_s = 0$ and neglecting the
tensor interaction) 
\bea
P_L &=& \frac{1}{\Delta} v \Bigg\{ 
\frac{4}{3 r} \lambda^2 m_B^6 \Big[ \vel B_2 \ver^2 - \vel D_2 \ver^2\Big] +
\frac{4}{r} \lambda m_B^2 m_\ell \, 
\mbox{\rm Re} [(B_1 - D_1) (B_4^\ast + B_5^\ast)] \nnb \\
&-& \frac{4}{r} \lambda m_B^4 m_\ell (1-r) \, 
\mbox{\rm Re} [(B_2 - D_2) (B_4^\ast + B_5^\ast)]+
\frac{32}{3} \lambda m_B^6 s \Big[ \vel A_1 \ver^2 - \vel C_1 \ver^2\Big] \nnb \\
&-&\frac{2}{r} \lambda m_B^4 s 
\Big[ \vel B_4 \ver^2 - \vel B_5 \ver^2\Big]+
\frac{4}{r} \lambda m_B^4 m_\ell s \, 
\mbox{\rm Re} [(B_3 - D_3) (B_4^\ast + B_5^\ast)] \nnb \\
&-&\frac{8}{3 r} \lambda m_B^4 (1-r-s)
\Big[ \mbox{\rm Re}(B_1 B_2^\ast) - \mbox{\rm Re}(D_1 D_2^\ast)\Big]
+\frac{4}{3 r} \lambda m_B^2 (\lambda + 12 r s) 
\Big[ \vel B_1 \ver^2 - \vel D_1 \ver^2\Big] \Bigg\}~, \nnb \\ \nnb \\
P_T &=& \frac{1}{\Delta} \sqrt{\lambda} \pi \Bigg\{ 
-8 m_B^3 m_\ell \sqrt{s} \, \mbox{\rm Re} [(A_1 + C_1) (B_1^\ast + D_1^\ast)] \nnb \\
&+& \frac{1}{2r} m_B^3 m_\ell (1+3 r + s) \sqrt{s} \, 
\Big[ 2 \mbox{\rm Re}(B_1 D_2^\ast) - 2 \mbox{\rm Re}(B_2 D_1^\ast)\Big] \nnb \\
&+&\frac{1}{r\sqrt{s}} m_B m_\ell (1- r - s)
\Big[ \vel B_1 \ver^2 - \vel D_1 \ver^2\Big] \nnb \\
&+&\frac{1}{r\sqrt{s}} m_B m_\ell^2 (1- r - s)
\Big[ 2 \mbox{\rm Re}(B_1 B_5^\ast) - 2 \mbox{\rm Re}(D_1 B_4^\ast)\Big] \nnb \\
&+&\frac{1}{2 r} m_B^3 m_\ell (1- r - s) \sqrt{s}\,
\mbox{\rm Re} [2 (B_1 + D_1) (B_3^\ast - D_3^\ast)] \nnb \\
&+&\frac{1}{r\sqrt{s}} m_B^3 m_\ell^2 \lambda 
\Big[ -2 \mbox{\rm Re}(B_2 B_5^\ast) + 2 \mbox{\rm Re}(D_2 B_4^\ast)\Big] \nnb \\ 
&+&\frac{1}{r\sqrt{s}} m_B^5 m_\ell(1-r) \lambda
\Big[ \vel B_2 \ver^2 - \vel D_2 \ver^2\Big]
+ \frac{1}{2r} m_B^5 m_\ell \lambda \sqrt{s} \,
\mbox{\rm Re} [-2 (B_2 + D_2) (B_3^\ast - D_3^\ast)] \nnb \\
&+&\frac{1}{2r\sqrt{s}} m_B^3 m_\ell [(1-r-s) ( 1-r) + \lambda ]
\Big[ -2 \mbox{\rm Re}(B_1 B_2^\ast) + 2 \mbox{\rm Re}(D_1 D_2^\ast)\Big] \nnb \\
&+&\frac{1}{2r\sqrt{s}} m_B (1-r-s)(-2 m_\ell^2  + m_B^2 s )
\Big[2 \mbox{\rm Re}(D_1 B_5^\ast) - 2 \mbox{\rm Re}(B_1 B_4^\ast)\Big] \nnb \\
&+&\frac{1}{2r\sqrt{s}} m_B^3 \lambda (-2 m_\ell^2  + m_B^2 s )
\Big[-2 \mbox{\rm Re}(D_2 B_5^\ast) + 2 \mbox{\rm Re}(B_2 B_4^\ast)\Big]
 \Bigg\}~, \nnb \\ \nnb \\ 
P_N &=& \frac{1}{\Delta} \pi v m_B^3 \sqrt{\lambda} \sqrt{s} \Bigg\{
8 m_\ell \, \mbox{\rm Im}(B_1^\ast C_1 + A_1^\ast D_1) \nnb \\
&+& \frac{1}{r} m_B^2 \lambda \Big[
\mbox{\rm Im}(m_\ell B_3-m_\ell D_3 - B_4) B_2^\ast +
\mbox{\rm Im}(-B_5-B_3 +D_3) D_2^\ast \Big] \nnb \\
&+& \frac{1}{r} m_\ell (1+3 r-s)\, 
\mbox{\rm Im} [(D_1 + B_1) (B_2^\ast - D_2^\ast)] \nnb \\
&+& \frac{1}{r} (1-r-s)\Big[ 
\mbox{\rm Im}(B_4 - m_\ell B_3+m_\ell D_3) B_1^\ast +
\mbox{\rm Im}(m_\ell B_3 - B_5-m_\ell D_3) D_1^\ast
\Big] \Bigg\}~,
\eea
where $\Delta$ is the expression within the curly parenthesis of the
unpolarized differential decay rate in Eq. (11).
These expressions for the longitudinal, transversal and normal polarizations
are general and model independent (if the tensor interaction is neglected).
It follows from the expressions of $P_T$ and $P_N$ that they are
proportional to the lepton mass and therefore they are nonvanishing only for
the $\tau$ lepton. 
In this work we also analyze the CP--violating asymmetry, which is defined as
\bea
A_{CP} (q^2) = \frac{\ds{\left(\frac{d \Gamma}{dq^2}\right)_{0} -
\left(\frac{d \overline \Gamma}{dq^2}\right)_{0}}}   
                 {\ds{\left(\frac{d \Gamma}{dq^2}\right)_{0}+
\left(\frac{d \overline \Gamma}{dq^2}\right)_{0}}}~, \nnb
\eea 
where $(d \Gamma/dq^2)_0$ is the unpolarized differential decay rate given by
Eq. (11) and $(d \overline \Gamma/dq^2)_0$ is the  unpolarized differential 
decay rate for the antiparticle channel. Note that in SM , CP--violating
asymmetry is equal to zero (or suppressed very strongly), since all form
factors are real (see below), Wilson coefficients $C_7^{eff}$ and $C_{10}$
are real and only $C_9^{eff}$ contains a strong phase. But this strong phase
can not lead to CP--violation itself.
Using these general expressions we can study the sensitivity of the
$\tau$--lepton polarizations on the new Wilson coefficients. Furthermore one
can investigate how strongly these polarizations deviate from the SM
predictions and for which Wilson coefficient this departure is more
essential.
But in the present work we will apply these general results to concrete
models, namely to the SM and three type of 2HDM, i.e., models I, II (about
models I and II, see for example \cite{R27} and III.
Note that in models I and II, the flavor
changing neutral currents which appear at tree level are avoided by imposing
ad hoc symmetry \cite{R28}. The phenomenological consequence of the 2HDM
without this discrete symmetry has been investigated in \cite{R29} (see also 
\cite{R30}--\cite{R39}). One novel feature of model III is existence of new
weak phase which appears in Yukawa interaction of fermions with Higgs fields
(see below). Existence of this new weak phase can lead to sizeable CP violation 
in $B \rar K^\ast \ell^+ \ell^-$ decay. Therefore if in future experiments
sizeable CP violation in the $B \rar K^\ast \ell^+ \ell^-$ decay is
discovered, it is an unambiguous indication of the existence of new physics
beyond SM, since in the SM the CP asymmetry suppressed very strongly.

Making the following replacements in the expressions given in Eq. (8), 
the explicit forms of $A_i$, $B_i$, $C_i$ and $D_i$  can be obtained in SM
and 2HDM easily.\\

{\bf 1.} SM

\bea
C_{LL}   &=& C_9^{eff}(m_b) - C_{10}(m_b)~,\nnb \\
C_{RL}   &=& 0~,\nnb \\
C_{BR}   &=& - 2 C_7^{eff}(m_b) ~,\nnb \\
C_{LR}   &=& C_9^{eff}(m_b) + C_{10}(m_b)~,\nnb \\
C_{RR}   &=& 0~,\nnb \\
C_{LRRL} &=& C_{RLLR} = C_{LRLR} = C_{RLRL} = 0~.
\eea

{\bf 2.} 2HDM

\bea
C_{LL}   &=& C_9^{eff\,2HDM}(m_b) - C_{10}^{2HDM}(m_b)~,\nnb \\
C_{RL}   &=& 0~,\nnb \\
C_{BR}   &=& - 2 C_7^{eff\,2HDM}(m_b) ~,\nnb \\           
C_{LR}   &=& C_9^{eff\,2HDM}(m_b) + C_{10}^{2HDM}(m_b)~,\nnb \\
C_{RR}   &=& 0~,\nnb \\
C_{LRRL} &=& C_{Q_1} ,\nnb \\
C_{RLRL} &=& C_{Q_2} ,\nnb \\
C_{LRLR} &=& C_{Q_1} ,\nnb \\
C_{RLLR} &=& - C_{Q_2}~.
\eea

The coefficients $C_i^{2HDM} (m_W)$ ($i=7,9$ and $10$) 
to the leading order are given by (see for example \cite{R40,R41})
\bea
C_7^{2HDM}(m_W) &=&
x \, \frac{(7-5 x - 8 x^2)}{24 (x-1)^3} +
\frac{x^2 (3 x - 2)}{4 (x-1)^4} \, \ln x \nnb \\
&+& \vel \lambda_{tt} \ver^2 \Bigg( \frac{y(7-5 y - 8 y^2)}
{72 (y-1)^3} + \frac{y^2 ( 3 y - 2)}{12 (y-1)^4} \, \ln y \Bigg) \nnb \\
&+& \lambda_{tt} \lambda_{bb} \Bigg( \frac{y(3-5 y)}{12 (y-1)^2} +
\frac{y (3 y - 2)}{6 (y-1)^3} \, \ln y \Bigg)~, \\ \nnb \\ \nnb \\
C_9^{2HDM}(m_W) &=& - \frac{1}{sin^2 \theta_W} \, B(m_W) +
\frac{1 - 4 sin^2 \theta_W}{sin^2 \theta_W} \, C(m_W) \nnb \\
&+& \frac{x^2(25-19 x)}{36 (x-1)^3} +
\frac{-3 x^4 + 30 x^3 - 54 x^2 + 32 x -8}{18 (x-1)^4} \, \ln x
+ \frac{4}{9} \nnb \\
&+& \vel \lambda_{tt} \ver^2 \Bigg[
\frac{1 - 4 sin^2 \theta_W}{sin^2 \theta_W} \, \frac{x y}{8} \Bigg(
\frac{1}{y-1} - \frac{1}{(y-1)^2} \, \ln y \Bigg)\nnb \\
&-& y \Bigg( \frac{47 y^2 - 79 y + 38}{108 (y-1)^3}
-\frac{3 y^3 - 6 y^3 + 4}{18 (y-1)^4} \, \ln y \Bigg) \Bigg]~,
\\ \nnb \\ \nnb \\
C_{10}^{2HDM}(m_W) &=& \frac{1}{sin^2 \theta_W} \Big( B(m_W) -
C(m_W) \Big) \nnb \\
&+& \vel \lambda_{tt} \ver^2 \frac{1}{sin^2 \theta_W} \,\frac{x y}{8}
\Bigg( - \frac{1}{y-1} + \frac{1}{(y-1)^2} \, \ln y \Bigg)~,\\ \nnb \\ \nnb \\
C_{Q_1} ( m_W ) &=& \frac{m_b m_\ell}{m_{h^0}^2} 
\frac{1}{\vel \lambda_{tt} \ver^2}
\frac{1}{sin^2 \theta_W} \frac{x}{4} \Bigg\{ \left(sin^2 \alpha + h\, cos^2
\alpha \right) f_1 (x,y) + \nnb \\
&+& \left[ \frac{m_{h^0}^2}{m_W^2} + \left(sin^2 \alpha + h\,
cos^2\alpha \right)(1-z) \right] f_2(x,y) +  \nnb \\
&+& \frac{sin^2 2 \alpha}{2 m_{H^\pm}^2} \left[m_{h^0}^2 -
\frac{(m_{h^0}^2 + m_{H^0}^2)^2}{2 m_{H^0}^2} \right] f_3 (y) \Bigg\}~,\\ \nnb \\ \nnb \\
C_{Q_2} (m_W) &=& \frac{m_b m_\ell}{m_{H^\pm}^2} \frac{1}{\vel \lambda_{tt} \ver^2} \left\{
f_1(x,y) +
\left[1+ \frac{m_{H^\pm}^2 - m_{A^0}^2}{m_W^2} \right] f_2(x,y) \right\}~,
\eea
where
\bea
x &=& \frac{m_t^2}{m_W^2}~,~~~~y=\frac{m_t^2}{m_{H^\pm}^2}~,
~~~~z=\frac{x}{y}~,~~~~h=\frac{m_{h^0}^2}{m_{H^0}^2}~, \nnb \\
B(x) &=& - \frac{x}{4 (x-1)} + \frac{x}{4 (x-1)^2} \, \ln x ~, \nnb \\
C(x) &=& \frac{x}{4} \Bigg( \frac{x-6}{2 (x-1)} +
\frac{3 x +2 }{2 (x-1)^2} \ln x \Bigg)~,\nnb \\
f_1 (x,y) &=& \frac{x\, \ln x}{x-1} - \frac{y\, \ln y}{y-1}~,\nnb \\
f_2(x,y) &=& \frac{x\, \ln y}{(z-x)(x-1)} + \frac{\ln z}{(z-1)(x-1)}~,\nnb \\
f_3(y) &=& \frac{1 -y + y\, \ln y}{(y-1)^2}~,
\eea
$sin^2\theta_W = 0.23$ is the Weinberg angle, $h^0,~H^0$ and $A^0$ are two
scalar and pseudoscalar Higgs fields, respectively. The coefficients 
$\lambda_{tt}$ and $\lambda_{bb}$ for model I and model II of the 2HDM are:
\bea
&&\lambda_{tt} = \cot \beta ~~,~~~~~ \lambda_{bb} = - \, \cot \beta
~~,~~~~~\mbox{\rm for model I}~, \nnb \\
&&\lambda_{tt} = \cot \beta ~~,~~~~~ \lambda_{bb} = + \, \tan  \beta
~~,~~~~~\mbox{\rm for model II}~,
\eea
while in model III $\lambda_{tt}$ or $\lambda_{bb}$ is complex, i.e.,
\bea
\lambda_{tt} \lambda_{bb} \equiv 
\vel \lambda_{tt} \lambda_{bb}\ver e^{i\phi}~.\nnb 
\eea
From Eqs. (16)--(20) we observe that the SM results for the Wilson
coefficients $C_7^{SM}(m_W)$, $C_9^{SM}(m_W)$ amd $C_{10}^{SM}(m_W)$ (and
correspondingly at $\mu=m_b$ scale) can all be obtained from 2HDM results by
making the following replacements
\bea
C_{Q_1} \rar 0 &,& C_{Q_2} \rar 0~,\nnb \\
C_7^{SM}(m_W) &=& C_7^{2HDM}(y \rar 0)~,\nnb \\
C_9^{SM}(m_W) &=& C_9^{2HDM}(y \rar 0)~,\nnb \\
C_{10}^{SM}(m_W) &=& C_{10}^{2HDM}(y \rar 0)~,\nnb
\eea

The evolution of the Wilson coefficients from the higher scale
$\mu = m_W$ down to the low energy scale $\mu = m_b$ is described by the
renormalization group equation.
The coefficients $C_7^{eff}(\mu),~C_9^{eff}(\mu),~C_{10}(\mu)$ at the scale
${\cal O}(\mu=m_b)$ are calculated in \cite{R42,R43} and $C_{Q_1}$ and 
$C_{Q_2}$ at the same scale to leading order are calculated in
\cite{R41}. The Wilson coefficient $C_{10}$ is not modified as
we move from $\mu=m_W$ to $\mu=m_b$ scale,
i.e., $C_{10}(m_b) \equiv C_{10}^{2HDM}(m_W)$.
In order to calculate $C_{9}^{2HDM}$ at $m_b$ scale, it is enough to make
the replacement $C_9^{SM}(m_W) \rar C_{9}^{2HDM}(m_W)$ and then
solve the corresponding renormalization group equation.
Hence, including the NLO
QCD corrections, $C_9^{eff}(m_b)$ can be written as:
\bea
\lefteqn{
C_9^{eff}(\mu) = C_9^{2HDM}(\mu)
\left[1 + \frac{\alpha_s(\mu)}{\pi} \omega (\hat s) \right]} \nnb \\
&&+ \, g(\hat m_c,\hat s) \Big[ 3 C_1(\mu) + C_2(\mu) + 3 C_3(\mu) +
C_4(\mu)
+ 3 C_5(\mu) + C_6(\mu) \Big] \nnb \\
&&- \frac{1}{2} g(0,\hat s) \ga C_3(\mu) + 3 C_4(\mu) \dr
-\,  \frac{1}{2} g \ga 1, \hat s\dr
\ga 4 C_3 + 4 C_4 + 3 C_5 + C_6 \dr \nnb \\
&&- \frac{1}{2} g \ga 0, \hat s\dr \ga C_3 + 3 C_4 \dr
+\, \frac{2}{9} \ga 3 C_3 + C_4 + 3 C_5 + C_6 \dr~,
\eea
where $\hat m_c = m_c/m_b~, ~\hat s = p^2/m_b^2$, and
\bea
\lefteqn{
\omega \ga \hat s \dr = - \frac{2}{9} \pi^2 -
\frac{4}{3} Li_2  \ga \hat s \dr - \frac{2}{3} \ln \ga \hat s\dr
\,\ln \ga 1 -\hat s \dr} \nnb \\
&&- \,\frac{5 + 4 \hat s}{3 \ga 1 + 2 \hat s \dr} \ln \ga 1 -\hat s \dr
-\frac{2 \hat s \ga 1 + \hat s \dr \ga 1 - 2 \hat s \dr}
{3 \ga 1 - \hat s \dr^2 \ga 1 + 2 \hat s \dr} \, \ln \ga \hat s\dr
+ \frac{5 + 9 \hat s - 6 {\hat s}^2}
{3 \ga 1 - \hat s \dr \ga 1 + 2 \hat s \dr}~
\eea
represents the ${\cal O}\ga \alpha_s \dr$ correction from the one gluon
exchange in the matrix element of $O_9$, while the function
$g \ga \hat m_c, \hat s \dr$ arises from one loop contributions of the
four--quark operators $O_1$--$O_6$, whose form is
\bea 
g \ga y_i, \hat s \dr &=& - \frac{8}{9}  \ln \ga \hat m_i\dr 
+ \frac{8}{27} + \frac{4}{9} y_i \nnb \\
&&- \frac{2}{9} \ga 2 + y_i \dr
 \sqrt{\vel 1-y_i \ver} \Bigg\{ \Theta \ga 1 - y_i \dr
\Bigg( \ln \frac{1+\sqrt{\vel 1-y_i \ver}}{1-\sqrt{\vel 1-y_i \ver}}
- i \, \pi \Bigg) \nnb \\
&&+ \Theta \ga y_i -1 \dr 2 \arctan \frac{1}{\sqrt{y_i - 1}} \Bigg\}~,
\eea
where $y_i = 4 {\hat m_i}^2/{\hat p}^2$.
The Wilson coefficient $C_9^{eff}$ receives also long distance contributions,  
which have their origin in the real $c\bar c$
intermediate states, i.e., $J/\psi$, $\psi^\prime$,
$\cdots$. The $J/\psi$ family is  
introduced by the Breit--Wigner distribution for the resonances
through the replacement \cite{R3,R6}
\bea
g \ga \hat m_c, \hat s \dr \rar g \ga \hat m_c, \hat s \dr -
\frac{3\pi}{\alpha^2_{em}} \, \kappa 
\sum_{V_i=J/\psi_i,\psi^\prime,\cdots}
\frac{m_{V_i} \Gamma(V_i \rar \ell^+ \ell^-)}
{(p^2 - m_{V_i}^2) + i m_{V_i} \Gamma_{V_i}}~,
\eea
where the phenomenological parameter $\kappa =2.3$ is chosen in order to
reproduce correctly the experimental value of the branching ratio
(see for example \cite{R44})

\section{Numerical analysis}

In this section we would like to present our numerical results. The main
free parameters $\lambda_{tt},~\lambda_{bb}$ of the 2HDM are restricted from 
$B \rar X_s \gamma$ decay, $B^0$--$\bar B^0$
mixing, $\rho$ parameter and neutron electric--dipole moment \cite{R36},
that yields $\vel \lambda_{bb} \ver = 50$, $\vel \lambda_{tt}\ver \le 0.03$.
Throughout the numerical analysis for the mass of the Higgs bosons we have used
$m_{h^0}=80~GeV,~m_{H^\pm}= 250~GeV,~m_{A^0}=250~GeV$ and $m_{H^0}=150~GeV$.
 
For the values of the form factors, we have used the results of
\cite{R25}, where  the radiative corrections to the leading twist
contribution and $SU(3)$ breaking effects are also taken into account.
The $q^2$ dependence of the form factors can be represented in terms of
three parameters as
\bea
F(q^2) = \frac{F(0)}{1-a_F\,\frac{q^2}{m_B^2} + b_F \left
    ( \frac{q^2}{m_B^2} \right)^2}~, \nnb
\eea
where, the values of parameters $F(0)$, $a_F$ and $b_F$ for the
$B \rar K^* \ell^+ \ell^-$ decay are listed in Table 1.

\begin{table}[h]                    
\renewcommand{\arraystretch}{1.5}                        
\addtolength{\arraycolsep}{3pt}
$$
\begin{array}{|l|ccc|}
\hline
& F(0) & a_F & b_F \\ \hline
A_1^{B \rar K^*} &
\phantom{-}0.34 \pm 0.05 & 0.60 & -0.023 \\
A_2^{B \rar K^*} &
\phantom{-}0.28 \pm 0.04 & 1.18 & \phantom{-}0.281\\
V^{B \rar K^*} &
 \phantom{-}0.46 \pm 0.07 & 1.55 & \phantom{-}0.575\\
T_1^{B \rar K^*} &
  \phantom{-}0.19 \pm 0.03 & 1.59 & \phantom{-}0.615\\
T_2^{B \rar K^*} & 
 \phantom{-}0.19 \pm 0.03 & 0.49 & -0.241\\
T_3^{B \rar K^*} & 
 \phantom{-}0.13 \pm 0.02 & 1.20 & \phantom{-}0.098\\ \hline
\end{array}   
$$
\caption{$B$ meson decay form factors in a three-parameter fit, where the
radiative corrections to the leading twist contribution and SU(3) breaking
effects are taken into account \cite{R25}.}
\renewcommand{\arraystretch}{1}
\addtolength{\arraycolsep}{-3pt}
\end{table}       

In Fig. (1) we present the dependence of the CP--violating asymmetry on $q^2$ 
and on the weak phase $\phi$ for the $B \rar K^\ast \tau^+ \tau^-$
decay in model III, since we have already noted that in SM and in models I
and II the CP--violating asymmetry is practically zero. 
We observe that CP asymmetry differs from zero in the
region $0 < \phi < 2 \pi$, except at $\phi=0,~\pi$ and $2 \pi$, and its
value in the region $0 < \phi < \pi$ ($\pi < \phi < 2 \pi$) is negative 
(positive).  

Fig .(2) depicts the dependence of the averaged CP asymmetry 
$\lla A_{CP} \rra$ (here and in all of the following discussions, by the
averaged values of the physical quantities we mean integration over $q^2$ in
the region $14~GeV^2 \le q^2 \le (m_B-m_{K^\ast})^2$) on the weak phase
angle $\phi$ in model III, taking into account short and long distance
contributions. It follows from this figure that 
$\lla A_{CP}\rra$ varies in the range (-0.04, 0.04) which is different from
zero and it definitely is an indication of the existence of new physics
beyond SM, since $\lla A_{CP} \rra$ is practically equal to zero in the SM.

In Fig. (3), the dependence of $P_L$ on $q^2$ and the weak phase angle $\phi$
without long distance effects in model III is presented. 
From this figure one can see that, for $q^2 > 14~GeV^2$,
$P_L$ varies in the range (-0.65,  -0.8) which is larger than the SM
prediction. This is due to the fact that the "new" contribution which comes
from the charged Higgs boson gives constructive interference to the SM
results. 

In Fig. (4) we present
the averaged longitudinal polarization $\lla P_L \rra$ on the weak phase
angle $\phi$, taking into account short and long distance contributions. 
For completeness the predictions of SM, model I and model II
on $\lla P_L \rra$ are also presented. It is observed from this figure that
$\lla P_L \rra$ in model III as modulo, is larger than the ones predicted by
SM, model I and model II. Therefore an observation of $\vel \lla P_L \rra
\ver \ge 0.65$ is another conclusive confirmation of the existence of new physics
beyond SM.
   
In Figs. (5) and (6) we present the dependence of the transversal and normal
polarizations of the $\tau$ lepton on $q^2$ and on the weak phase angle
$\phi$, respectively, without long distance effects in model III. 
The dependence of the averaged transversal and normal
polarizations on the weak phase angle, taking into account short and long
distance contributions, are depicted in Figs. (7) and (8),
respectively. For sake of completeness we presented also the predictions
of SM, model I and model II of the same physical quantity in both figures.
Fig. (7) clearly depicts that, the prediction of model III on $\lla P_T \rra$ 
as modulo, is approximately five times smaller than the ones predicted by SM, 
model I and model II. However the situation is totally different for 
$\lla P_N \rra$, having a range of values in the region (0.10, 0.15) in
model III, it is approximately two or three times larger than the ones
predicted by SM, model I and model II. 

Here we would like to make the following remark. It follows from Eq. (13)
that $P_N$ is defined as the imaginary part of the form factors and of the
corresponding Wilson coefficients $C_7^{eff},~C_9^{eff}$, $C_{10}$, $C_{Q_1}$ and 
$C_{Q_2}$.  In
SM  $C_7^{eff}$ and $C_{10}$ are real and only $C_9^{eff}$ has imaginary part. On
the other side all theoretical methods predict these form factors to be real
quantities. For this reason, if in future experiments a different value for 
$P_N$ were observed compared to the SM prediction, it is an indication of
unambiguous information about the existence of the above--mentioned
CP--violating phase in theory. 

In conclusion, we have investigated the exclusive $B \rar K^\ast \tau^+ \tau^-$
decay in the SM and in three different versions of the 2HDM. From the
results we have obtained we conclude that the combined analysis of the
CP--violating asymmetry and $\tau$ lepton polarization effects are very
useful tools in looking for new physics beyond SM.

\newpage

\newpage
\section*{Figure captions}
{\bf Fig. 1} The dependence of the CP--violating asymmetry $A_{CP}$
of the $\tau$ lepton on $q^2$ and on the weak phase $\phi$ in model III.\\ \\
{\bf Fig. 2} The dependence of the averaged CP asymmetry $\lla A_{CP} \rra$
of the $\tau$ lepton on the weak phase $\phi$ in model III, taking into
account short and long distance contributions.\\ \\
{\bf Fig. 3} The dependence of the longitudinal polarization $P_L$
of the $\tau$ lepton on $q^2$ and on the weak phase $\phi$ in model III,
taking into account only the short distance contribution in $C_9^{eff}$.\\ \\
{\bf Fig. 4} The dependence of the averaged longitudinal polarization
$\lla P_L \rra$ of $\tau$ lepton on the weak phase $\phi$, taking into
account short and long distance contributions. \\ \\
{\bf Fig. 5} The dependence of the transversal polarization $P_T$
of the $\tau$ lepton on $q^2$ and on the weak phase $\phi$ in model III,
taking into account only the short distance contribution in $C_9^{eff}$.\\ \\ 
{\bf Fig. 6} The dependence of the normal polarization $P_N$  
of the $\tau$ lepton on $q^2$ and on the weak phase $\phi$ in model III,
taking into account only the short distance contribution in $C_9^{eff}$.\\ \\
{\bf Fig. 7} The dependence of the averaged transversal polarization $\lla P_T \rra$
of the $\tau$ lepton on the weak phase $\phi$, taking into
account short and long distance contributions.\\ \\       
{\bf Fig. 8} The dependence of the averaged normal polarization $\lla P_N \rra$
of the $\tau$ lepton on the weak phase $\phi$, taking into
account short and long distance contributions.\\ \\       

\newpage

\begin{figure}[H]
\vskip 1.5 cm
    \includegraphics{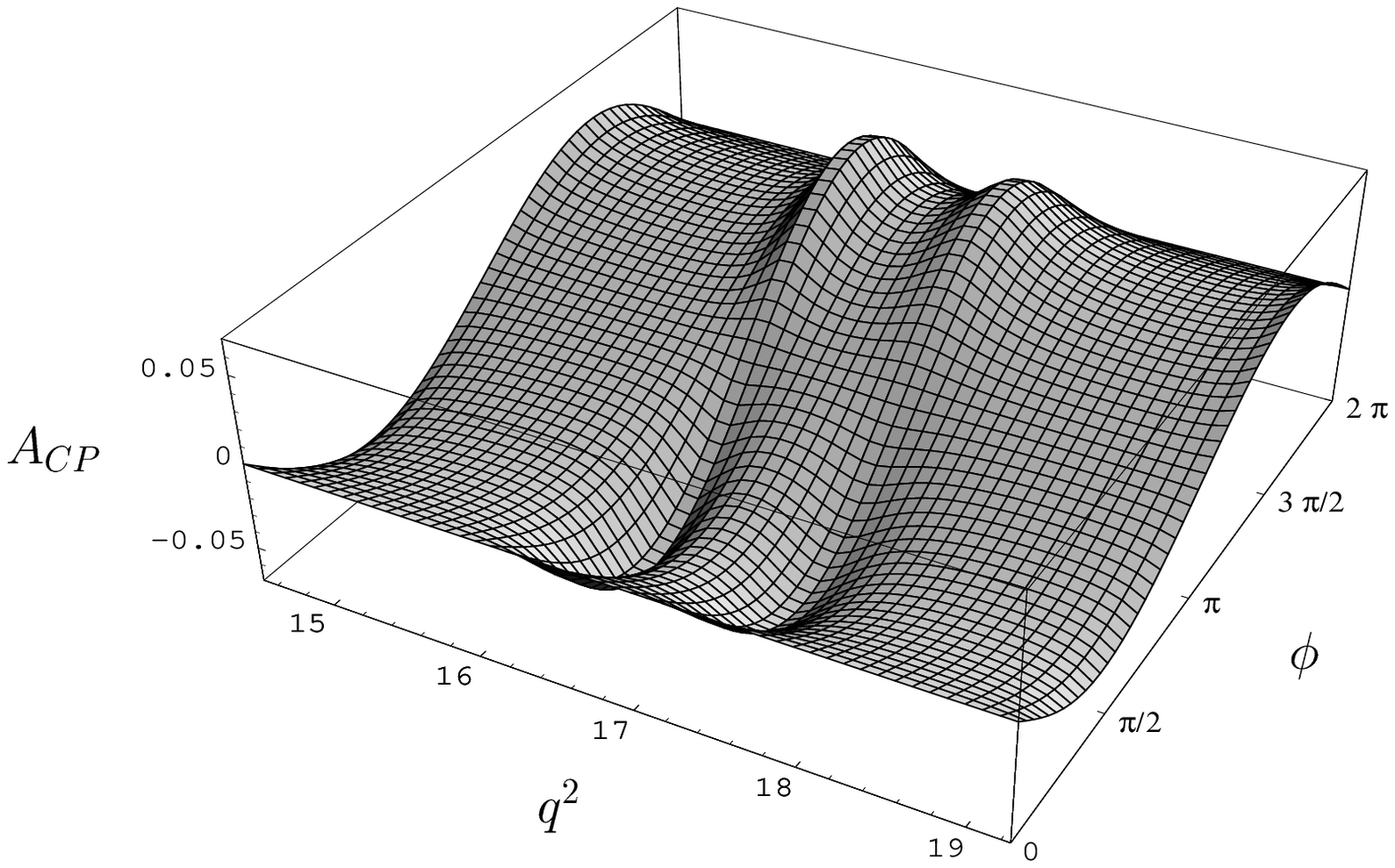}
\vskip 6.2cm   
\caption{}
\end{figure}

\begin{figure}
\vskip 1.5 cm
    \includegraphics{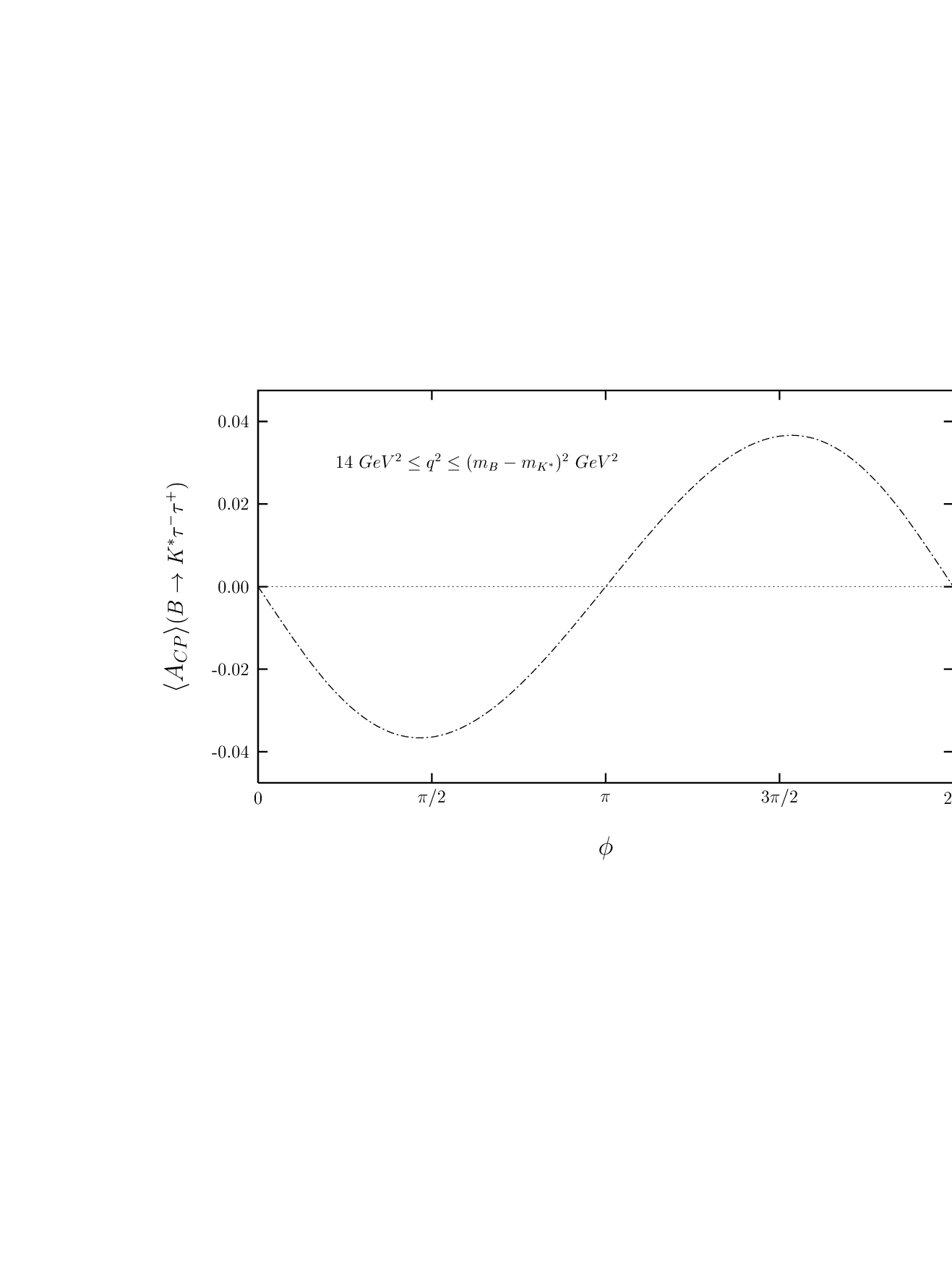}
\vskip 7.9 cm
\caption{}
\end{figure}

\begin{figure}
\vskip 1.5 cm
    \includegraphics{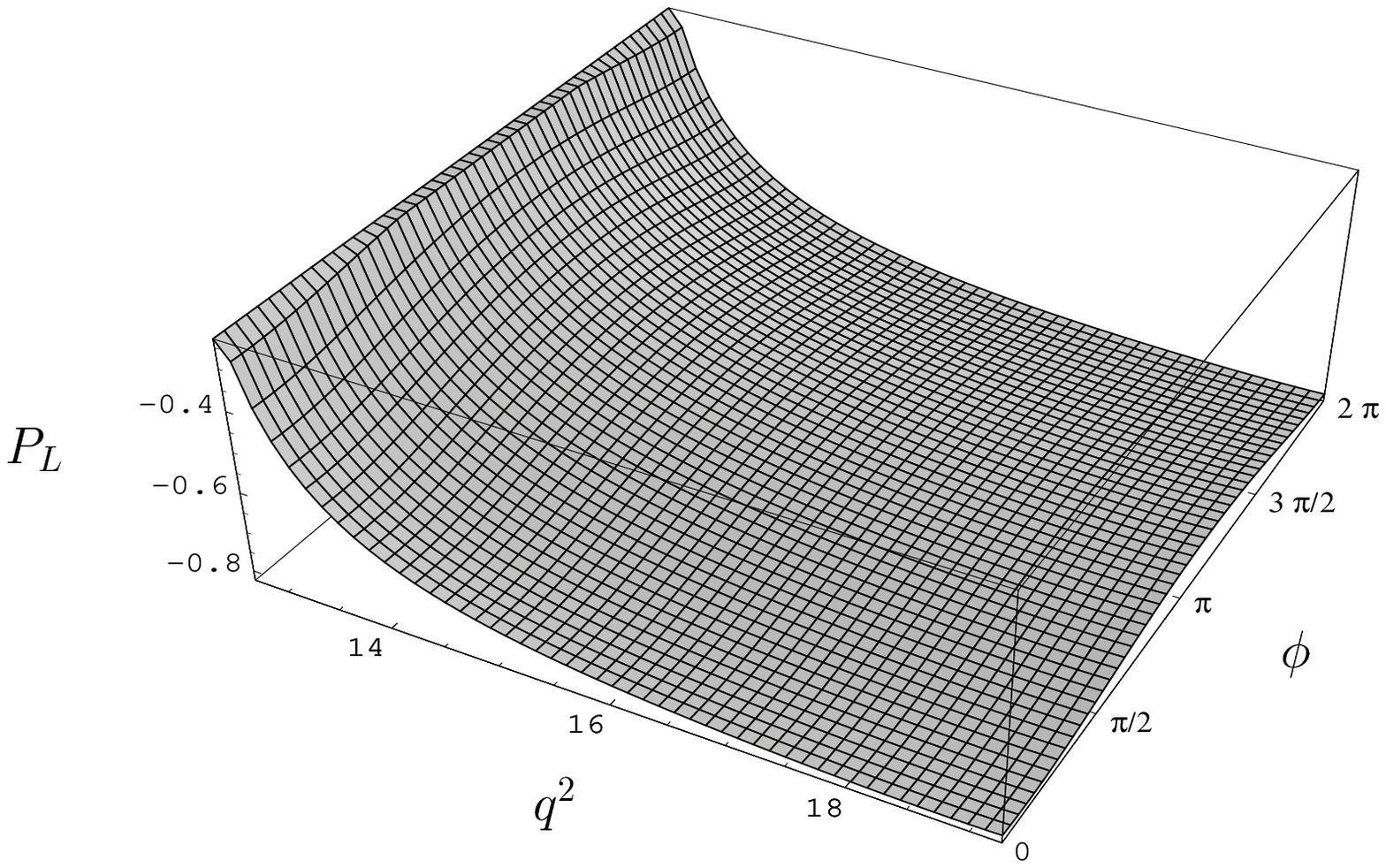}
\vskip 6.2cm   
\caption{}
\end{figure}

\begin{figure}
\vskip 1.5 cm
    \includegraphics{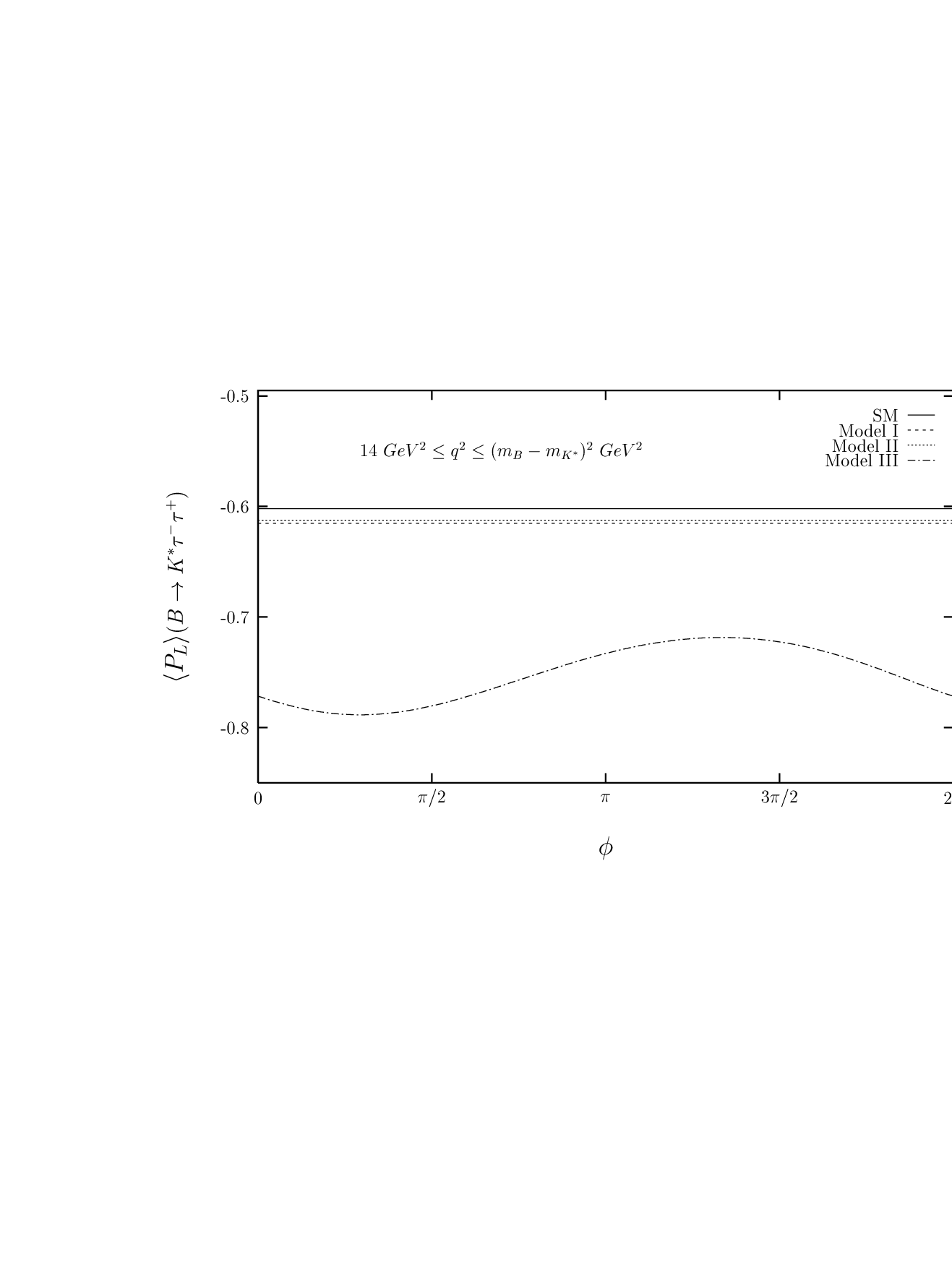}
\vskip 7.9 cm
\caption{}
\end{figure}

\begin{figure}
\vskip 1.5 cm
    \includegraphics{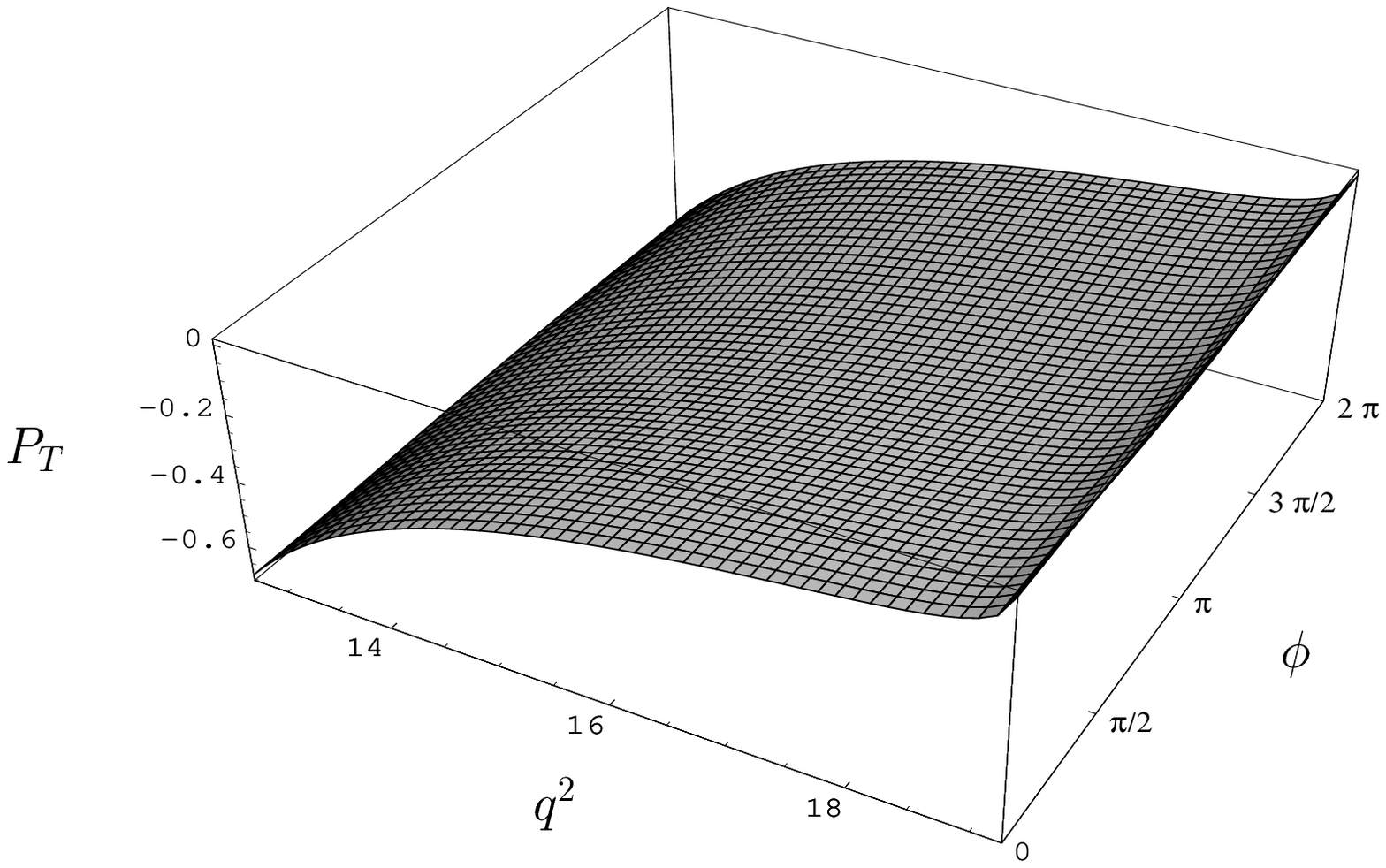}
\vskip 6.2cm   
\caption{}
\end{figure}

\begin{figure}
\vskip 1.5 cm
    \includegraphics{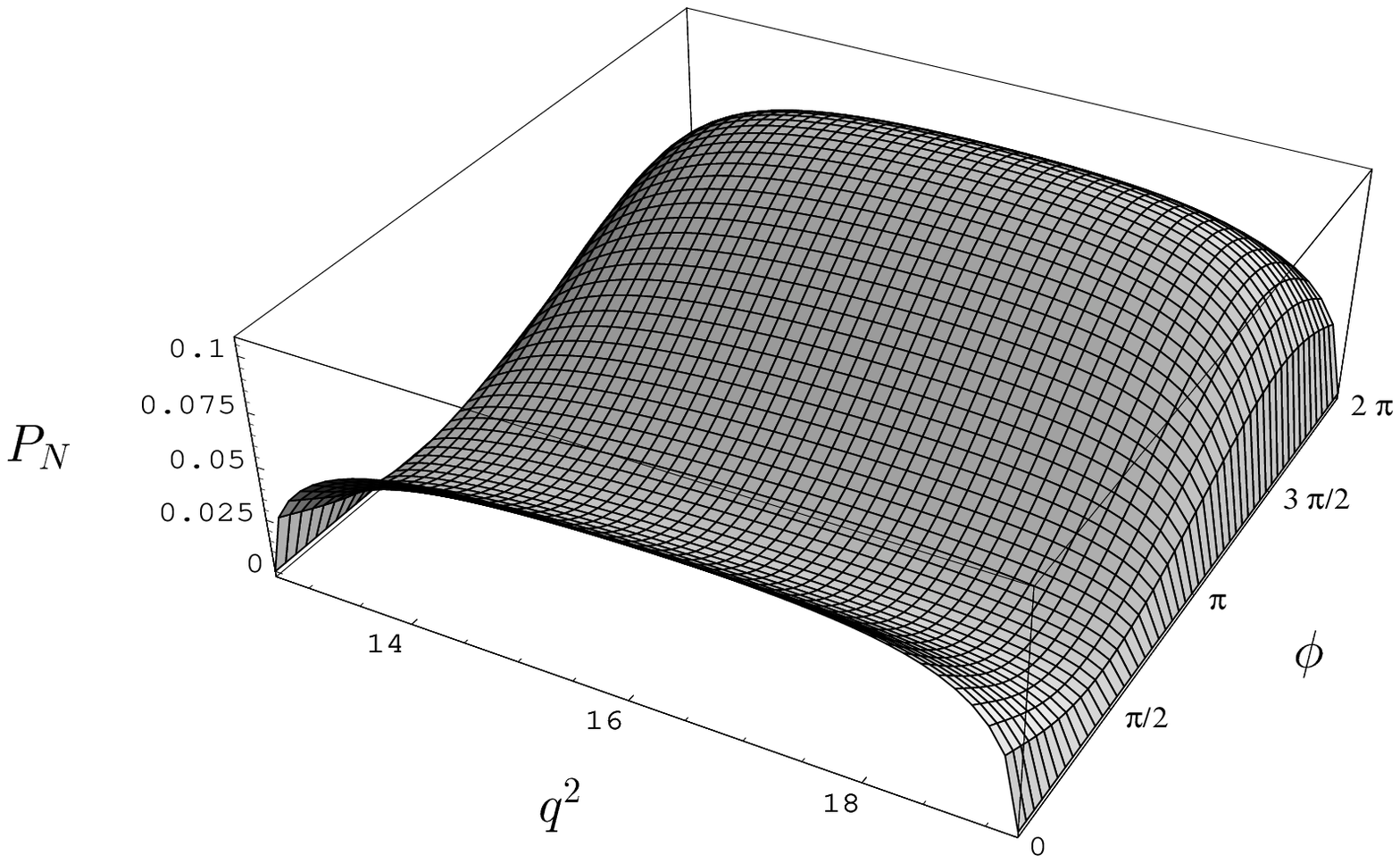}
\vskip 6.2 cm
\caption{}
\end{figure}

\begin{figure}
\vskip 1.5 cm
    \includegraphics{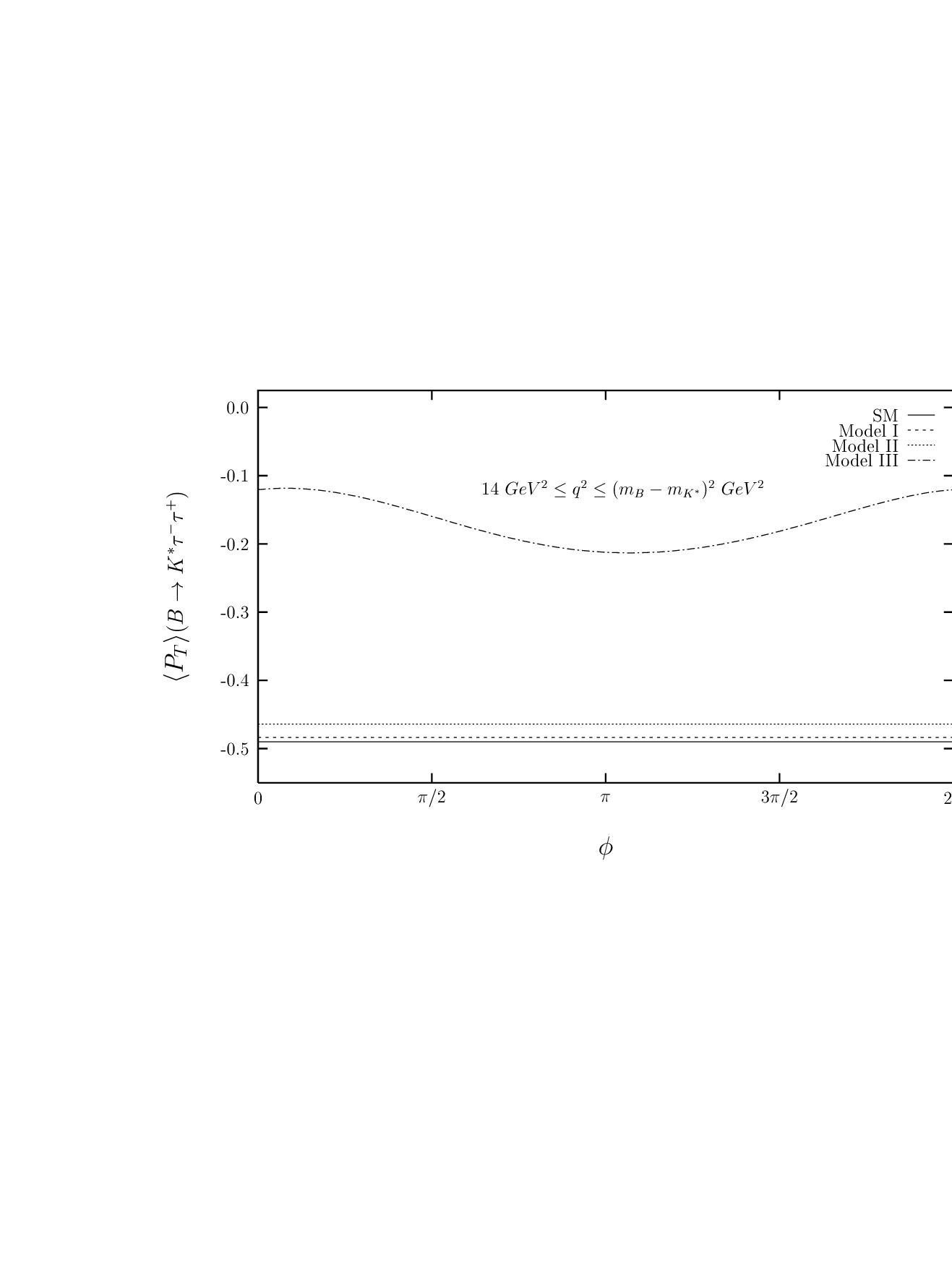}
\vskip 7.1cm   
\caption{}
\end{figure}

\begin{figure}
\vskip 1.5 cm
    \includegraphics{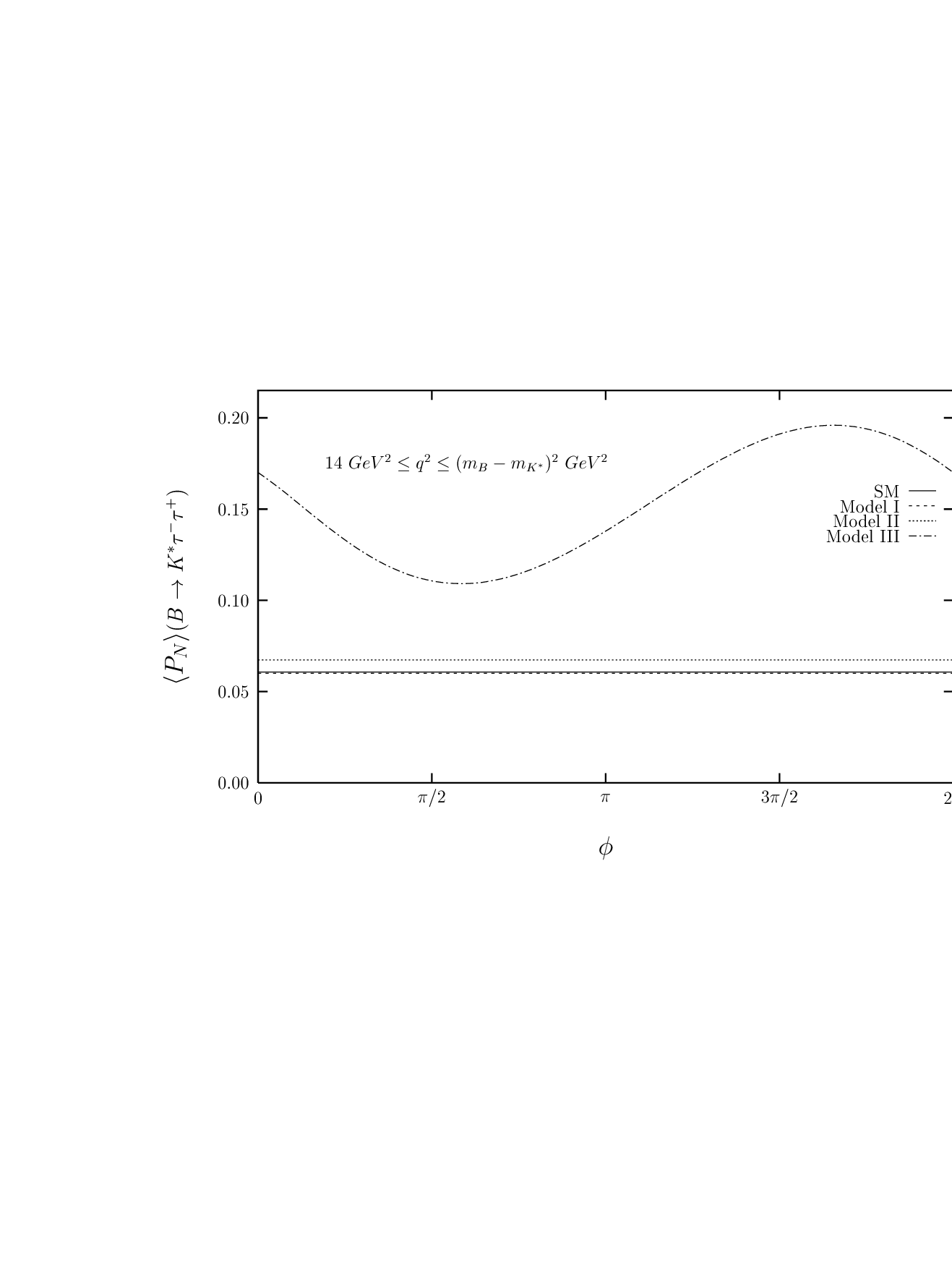}
\vskip 7.9 cm
\caption{}
\end{figure}

\end{document}